\ifdefined\PRBPREPRINT
  \documentclass[aps,prb,preprint,superscriptaddress,longbibliography,floatfix]{revtex4-2}
\else
  \documentclass[aps,prb,reprint,superscriptaddress,longbibliography,floatfix]{revtex4-2}
\fi

\usepackage[T1]{fontenc}
\usepackage[utf8]{inputenc}
\usepackage{amsmath,amssymb}
\usepackage{graphicx}

\begin{document}

\title{Magnetoplasmon molecule}

\author{A. V. Larionov}
\affiliation{Institute of Solid State Physics named after Yu. A. Osipyan, Russian Academy of Sciences, 142432 Chernogolovka, Moscow District, Russia}

\author{A. V. Gorbunov}
\affiliation{Institute of Solid State Physics named after Yu. A. Osipyan, Russian Academy of Sciences, 142432 Chernogolovka, Moscow District, Russia}

\author{A. B. Van'kov}
\affiliation{Institute of Solid State Physics named after Yu. A. Osipyan, Russian Academy of Sciences, 142432 Chernogolovka, Moscow District, Russia}

\author{D. A. Shchigarev}
\email{dima.shigarev@yandex.ru}
\affiliation{Institute of Solid State Physics named after Yu. A. Osipyan, Russian Academy of Sciences, 142432 Chernogolovka, Moscow District, Russia}
\affiliation{Moscow Institute of Physics and Technology, Dolgoprudny, Moscow Region, Russia}
\affiliation{Skolkovo Institute of Science and Technology, 121205 Moscow, Russia}

\author{L. V. Kulik}
\affiliation{Institute of Solid State Physics named after Yu. A. Osipyan, Russian Academy of Sciences, 142432 Chernogolovka, Moscow District, Russia}

\author{V. Umansky}
\affiliation{Braun Center for Submicron Research, Weizmann Institute of Science, 76100 Rehovot, Israel}

\begin{abstract}
We investigate a two-dimensional electron system in an external magnetic field. An intrinsic four-particle excitation of the electron system, which we term a magnetoplasmon molecule, is observed. The molecule comprises two identical electrons and two Fermi holes with opposite spins. Both electrons and one Fermi hole participate in the collective plasma oscillations of the electron system, whereas the second Fermi hole neutralizes the excess electronic charge. We examine the interaction between the magnetoplasmon molecule and a magnetoexciton condensate.
\end{abstract}

\maketitle

\section{Introduction}

A theoretical study showed that a plasmon molecule---a bound state of two plasmons in a metal---is not formed in a metallic electron gas \cite{dubois1978}. More recent studies of high-mobility semiconductor and graphene systems demonstrated the formation of a three-particle quasiparticle comprising a plasmon bound to an additional charge carrier, namely, a plasmaron \cite{bostwick2010,zhuravlev2016,dial2012}. Two-particle bound plasmon-exciton states known as plexcitons have also attracted considerable attention \cite{fofang2011,balci2015,torma2015,zhao2025}. In metallic nanoparticles, quantization of the surface-plasmon energy enables the strong-coupling regime: the electric field of a surface plasmon wave in the metal is resonant with excitonic transitions in the surrounding organic-semiconductor matrix, producing plexcitons. Analogous magnetoplasmon-phonon states have been reported in semiconductor two-dimensional systems subjected to a quantizing magnetic field \cite{kulik2000}. The electron system then becomes quasi-zero-dimensional, and tuning the magnetic-field quantization energy brings the magnetoplasmon energy into resonance with the optical-phonon energy of the host heterostructure. This establishes strong coupling between magnetoplasmons and optical phonons. In all of these examples, a single plasmon interacts with additional charged or neutral quasiparticles, producing new two- or three-particle plasma states. No experimental evidence has yet been reported for a four-particle molecular complex formed from two plasmons. Here we identify the conditions under which a hybrid magnetoplasmon molecule forms; this state was first observed in Ref.~\onlinecite{zhuravlev2016}. The molecule is hybrid because, although it comprises two electrons and two Fermi holes, both electrons but only one Fermi hole participate in the plasma oscillations. The second Fermi hole, whose spin is opposite to that of the first, neutralizes the excess electron charge and renders the molecule electrically neutral, but does not participate in the plasma oscillations. A molecule of this type has not been considered in electron-gas theory. Nevertheless, under appropriate conditions it becomes a well-defined four-particle excitation of an electron system in a magnetic field.

Our experimental approach to observing three- and four-particle bound magnetoplasmon states in GaAs/AlGaAs quantum wells is based on probing the excitation spectrum of a quantum Hall insulator in the presence of a dense ensemble of quasi-equilibrium two-particle excitations. At electron filling factor $\nu$ = 2, the lowest-energy two-particle excitations of the insulating ground state are cyclotron magnetoexcitons, each formed by an excited electron in the empty first Landau level and a Fermi hole in the completely filled zeroth Landau level. The two-particle spectrum contains two types of cyclotron magnetoexcitons: a spin-singlet magnetoexciton with total spin $S$ = 0 and a spin-triplet magnetoexciton with total spin $S$ = 1. The spin-singlet magnetoexciton is a magnetoplasmon, i.e., an excitation associated with collective plasma oscillations of the electron system \cite{bychkov1981,kallin1984}. The three components of the spin-triplet cyclotron exciton, with spin projections -1, 0, and +1 along the magnetic field, are split by the Zeeman energy. At zero generalized momentum, the energy of the lowest triplet component lies below the cyclotron energy by an amount determined by the Zeeman energy and Coulomb correlations \cite{kulik2005}. This cyclotron spin-flip exciton (CSFE) is therefore the lowest-energy excitation of the quantum Hall insulator. Unlike the magnetoplasmon, it is optically inactive: the CSFE is a dark exciton that does not couple to the electromagnetic field in the dipole approximation. Nevertheless, high-energy dipole-allowed interband optical transitions in a heterostructure containing a two-dimensional electron system (2DES) can create a dense quasi-equilibrium ensemble of CSFEs \cite{kulik2015}.

\begin{figure*}[t]
\includegraphics[width=\textwidth]{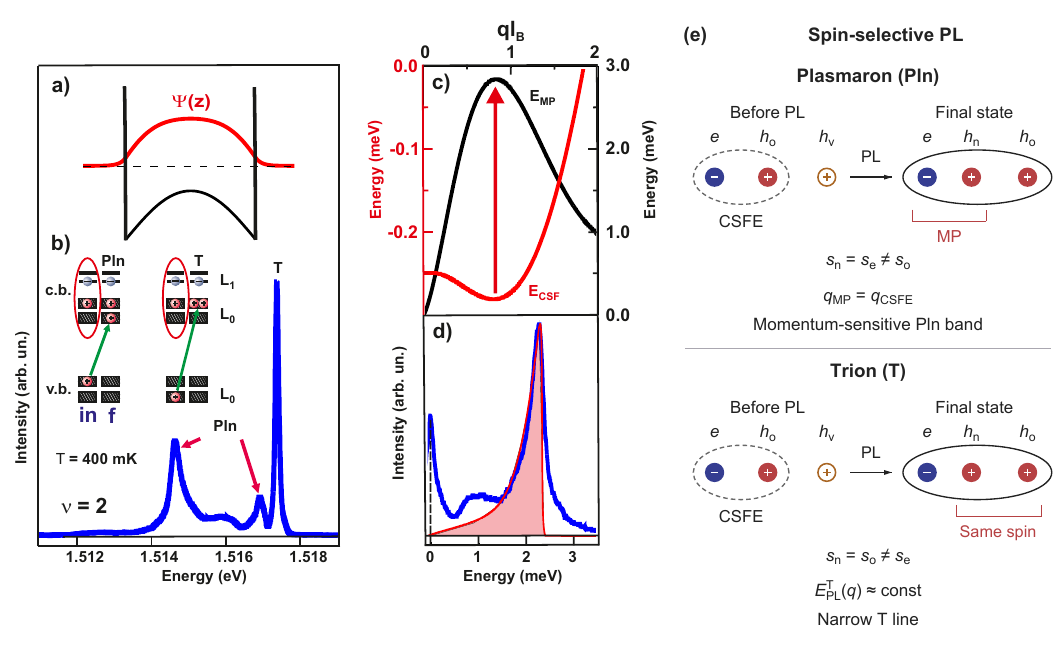}
\caption{(a) Calculated quantum-well potential and lowest-subband envelope wave function. (b) PL spectrum at $\nu=2$ and $T=400~\mathrm{mK}$, with $d_{\mathrm{pump}}\approx0.8~\mathrm{mm}$ and $I_{\mathrm{pump}}\approx2~\mathrm{mW\,cm^{-2}}$; the insets show the plasmaron (Pln) and trion (T) channels. (c) Calculated CSFE and magnetoplasmon dispersions. (d) Energy-loss spectrum of the photoexcited-hole recombination; the red curve is calculated for a Boltzmann CSFE distribution. (e) Spin-selective PL pathways. Recombination of $h_v$ with $e_0$ creates $h_n$. In the Pln channel, $e$ and $h_n$ form the magnetoplasmon component and $q_{\mathrm{MP}}=q_{\mathrm{CSFE}}$; in the T channel, $h_n$ and $h_o$ occupy the same spin sublevel. Arrows denote optical transitions; positions are schematic.}
\label{fig:fig1}
\end{figure*}

CSFEs are excited as follows. The electron system is photoexcited at photon energies well above the band gap of the GaAs/AlGaAs quantum well. The high-energy photoexcited electrons relax to the empty first Landau level of the conduction band, whereas the photoexcited holes relax to the zeroth heavy-hole Landau level of the valence band. During relaxation, a valence-band hole can flip its spin several times because of the strong spin-orbit interaction in the quantum-well valence band. It subsequently recombines with an equilibrium conduction-band electron, leaving a vacancy (a Fermi hole) in the zeroth electron Landau level. If the photoexcited hole changes its spin projection by one during valence-band relaxation, its conversion into a Fermi hole changes the spin projection of the entire electron system by one. By contrast, the spin of the photoexcited electron remains unchanged during relaxation because spin-orbit coupling in the GaAs conduction band is weak. Because radiative relaxation of a CSFE to the ground state is forbidden, these excitations have exceptionally long lifetimes \cite{dickmann2013}. Relaxation is further inhibited because the minimum of the CSFE dispersion occurs not at zero generalized momentum but near the inverse magnetic length, at $q\sim1/l_B$, where $l_B=\sqrt{\hbar c/(eB)}$ \cite{kallin1984} [Fig.~\ref{fig:fig1}(c)]. Consequently, relaxation of the lowest-energy CSFEs requires the transfer of both energy and a large momentum to other quasiparticles. CSFE lifetimes can therefore reach the millisecond range \cite{kulik2016}. Such long lifetimes allow a weak continuous-wave photoexcitation, which does not overheat the electron system, to generate quasi-equilibrium CSFE ensembles with densities of approximately $10^{10}~\mathrm{cm}^{-2}$.

If an additional electron-hole pair is excited in this system containing a dense quasi-equilibrium CSFE ensemble, the photoexcited valence-band hole recombines with an equilibrium conduction-band electron. The resulting Fermi hole can bind to a CSFE to form either a trion, denoted T, or a plasmaron, denoted Pln, depending on the spin polarization of the Fermi hole [see the schemes at the top of Fig.~\ref{fig:fig1}] \cite{zhuravlev2016}. The trion is a three-particle excitation comprising a CSFE and the Fermi hole created by recombination of the photoexcited valence-band hole (Fig.~\ref{fig:fig1}). The newly formed Fermi hole has the same spin as the Fermi hole in the CSFE. Because the trion energy depends only weakly on the momentum of the parent CSFE, the trion appears as a narrow line in the photoluminescence (PL) spectrum. The plasmaron band is an energy band of three-particle states produced when a valence-band hole recombines with an equilibrium electron in the lower spin sublevel of the zeroth Landau level.

In this case, the newly formed Fermi hole has the same spin as the CSFE electron and the opposite spin to the Fermi hole within the CSFE. The intrinsic excited state formed by this Fermi hole and the CSFE is a plasmaron: a magnetoplasmon bound to the Fermi hole that belonged to the CSFE before recombination. Plasmaron-producing recombination obeys an experimentally established conservation rule: the momentum of the initial CSFE is transferred to the magnetoplasmon formed in the final state by the newly created Fermi hole and the CSFE electron [Fig.~\ref{fig:fig1}(c)] \cite{kuznetsov2018}. The energy of the optical transition that produces a plasmaron is therefore reduced by the energy of a magnetoplasmon with the momentum of the parent CSFE (Fig.~\ref{fig:fig1}). Accordingly, the plasmaron energy distribution can be interpreted as the CSFE momentum distribution immediately before recombination [Fig.~\ref{fig:fig1}(d)]. The other Fermi hole from the CSFE remains bound to the magnetoplasmon, reducing the magnetoplasmon contribution to the plasmaron energy by approximately one fifth relative to the theoretical energy of an isolated magnetoplasmon [Fig.~\ref{fig:fig1}(c)]. Accounting for this reduction yields a theoretical plasmaron spectrum for a Boltzmann distribution of gas-phase CSFEs [Fig.~\ref{fig:fig1}(d)], which can be compared with the energy-loss spectrum for valence-hole recombination in the excited electron system relative to recombination in the unexcited system. The Boltzmann distribution describes the spectrum reasonably well, although a fraction of the gas-phase CSFEs evidently remains out of equilibrium under the continuous-wave excitation.

It is worth noting that a neutral photon cannot directly create a charged three-particle electronic excitation. Absorbed photons create neutral electron-hole pairs, whose electron and hole are initially generated at the same point in space because the photon momentum is small. During relaxation from high-energy states, however, coupling to the heterostructure phonon subsystem can spatially separate the electron and hole. Spatially separated electrons and holes can each be captured by a CSFE, producing electron and hole trions that are far apart; their mutual Coulomb interaction can then be neglected. Subsequent recombination converts the photoexcited valence-band hole into a Fermi hole and produces either a trion or a plasmaron. If, instead, the additional Fermi hole and the photoexcited electron remain spatially co-located during photoexcitation and relaxation, the initial state can be a biexciton composed of a CSFE and the photoexcited electron-hole pair. Recombination of the valence-band hole then leaves either two CSFEs or one CSFE and one magnetoplasmon. Two CSFEs do not form a bound four-particle complex because their interaction is repulsive \cite{dickmann2019}. A hybrid molecular state comprising a CSFE and a magnetoplasmon, however, is not theoretically forbidden. In that state, the two electrons occupy the same spin sublevel of a Landau level and are therefore indistinguishable. Both electrons can participate in collective plasma oscillations with the newly created Fermi hole. Direct calculations of the excitation spectrum for systems of 20 and 30 electrons reveal a vast number of intrinsic four-particle excited states with identical spin quantum numbers but different orbital quantum numbers. It is not known a priori which of these states are bound molecular states. The existence of a hybrid magnetoplasmon molecule must therefore be established experimentally.

\section{Experiment}

We investigated a high-quality heterostructure containing a single symmetrically doped 31-nm-wide GaAs/AlGaAs quantum well (QW), with a 2D-channel electron density of $n_e=2\times10^{11}~\mathrm{cm}^{-2}$ and a dark mobility of $\mu_e=1.5\times10^7~\mathrm{cm}^2/(\mathrm{V\,s})$. A sample approximately $3\times3~\mathrm{mm}^2$ in area was mounted in a helium cryostat equipped with a superconducting solenoid; the magnetic field was perpendicular to the QW plane.

Spectroscopic measurements were performed in two configurations: with optical fibers in a dilution refrigerator and with an optical system that focused the excitation light through the window of a pumped ${}^{3}\mathrm{He}$ cryostat. For measurements between 40 and 650 mK in the dilution refrigerator, we used a pair of multimode fused silica fibers. One fiber delivered the pump-laser radiation to the sample, whereas the other collected the photoluminescence and guided it to the entrance slit of a grating spectrometer equipped with a cooled charge-coupled device (CCD) camera. The fibers had a diameter of 200 $\mu\mathrm{m}$ and a numerical aperture of 0.22, yielding a pump-spot diameter of $d_{\mathrm{pump}}\simeq0.8~\mathrm{mm}$ on the sample. A continuous-wave single-mode diode laser (wavelength $\lambda\simeq785~\mathrm{nm}$) generated the nonequilibrium CSFE ensemble by nonresonant, subbarrier photoexcitation, for which the pump-photon energy was below the band gap of the heterostructure barrier.

Measurements with tight focusing of the pump beam employed a ${}^{3}\mathrm{He}$ insert with an optical window, mounted inside a ${}^{4}\mathrm{He}$ cryostat equipped with a superconducting solenoid. These experiments were performed at 0.55--1.5 K. A collimated beam from a continuous-wave single-mode diode laser ($\lambda\approx785~\mathrm{nm}$) was used both to generate the nonequilibrium CSFE ensemble and to excite the PL signal.

A prerequisite for studying four-particle bound states is that the density of high-energy photoexcited electron-hole pairs be much lower than that of the CSFE ensemble. The density ratio is set by the ratio between the recombination time of a photoexcited valence-band hole, approximately 100 ps, and the CSFE lifetime. In our experiment, a gas-phase CSFE has a lifetime of approximately 100 $\mu\mathrm{s}$, six orders of magnitude longer than the valence-hole recombination time. The CSFE density therefore exceeds the density of photoexcited electron-hole pairs by the same factor. The properties of four-particle excited states are consequently measured in the single-excitation limit, in which interactions between such excitations can be neglected.

Increasing the probability that a photoexcited electron-hole pair and a CSFE form a four-particle state at the same spatial location requires a higher local density of photoexcited pairs in the presence of a dense CSFE ensemble. Increasing the total pump power $P_{\mathrm{pump}}$, however, heats the electron system and destroys its many-particle states. We therefore varied the excitation-spot size while keeping the total pump power low. The spot size was chosen so that the formation dynamics of the hybrid magnetoplasmon molecule could be tracked as the pump power density $I_{\mathrm{pump}}$ was varied over two orders of magnitude.

A high-numerical-aperture two-lens imaging system was installed inside the ${}^{3}\mathrm{He}$ insert to focus the pump radiation onto the sample surface. Precise focusing was achieved by translating the sample smoothly along the optical axis with a mechanical feedthrough. The pump-spot diameter on the sample was $d_{\mathrm{pump}}\approx10$ $\mu\mathrm{m}$. The same lens pair collected the PL emission and collimated it outside the cryostat. A long-focal-length objective projected a magnified image of the sample onto the entrance slit of a grating spectrometer equipped with a cooled CCD camera. The PL image was recorded in the zeroth diffraction order. Spectral selection was provided by a 10-nm-bandwidth interference filter, and the spatial resolution was as high as approximately 1 $\mu\mathrm{m}$. Reflection from the sample surface was suppressed by crossed linear polarizers placed outside the cryostat, one in the excitation path and the other in the collection path.

In the fiber-coupled dilution-refrigerator measurements, the pump power at the sample was limited to $P_{\mathrm{pump}}$ $\leq$ 10 $\mu\mathrm{W}$ to avoid overheating, corresponding to $I_{\mathrm{pump}}$ $\approx$ 2 mW cm\textsuperscript{-2}. In the spatially resolved measurements with the optical ${}^{3}\mathrm{He}$ insert, the useful ranges were 0.1--20 $\mu\mathrm{W}$ in pump power and 0.1--20 W cm\textsuperscript{-2} in pump power density. This configuration therefore provided much stronger local photoexcitation at a substantially lower total pump power.

The parameters of the dark-CSFE ensemble were monitored through resonant reflection at the ``0-0'' optical transition from the zeroth heavy-hole Landau level in the valence band to the zeroth electron Landau level in the conduction band, thereby detecting the nonequilibrium Fermi holes contained in the CSFEs. This reflection is absent at equilibrium because the zeroth electron Landau level is completely filled. It appears only when long-lived CSFEs are generated in the 2DES and is termed photoinduced resonant reflection (PRR) \cite{kulik2015}. At $T\lesssim1~\mathrm{K}$ and a CSFE concentration $N_{\mathrm{ex}}\sim(1\text{--}10)\%$ of the magnetic-flux-quantum density $N_{\Phi}$, lowering the temperature at fixed pump power density $I_{\mathrm{pump}}$ drives the excited electron system into a condensed state, the magnetofermionic condensate \cite{kulik2016}. Condensation occurs in a quasi-equilibrium excited system of 2D fermions: at equilibrium they fill the zeroth Landau level, while photoexcitation promotes a fraction to the empty first Landau level. The resulting fermionic system contains a gas of CSFEs, which are composite bosons; below a critical temperature this gas enters a condensed state. Because the condensing magnetoexcitons are CSFEs, this state may also be called a magnetoexciton condensate. It exemplifies condensation in generalized momentum space, whose coordinates depend on both spatial coordinates and their gradients \cite{avron1978}.

A dense steady-state CSFE ensemble can form only in selected GaAs/AlGaAs heterostructures. The principal experimental requirement is a CSFE lifetime long enough for these quasiparticles to accumulate under weak continuous-wave photoexcitation. This condition is difficult to satisfy. If random-potential broadening causes the spin sublevels of the zeroth electron Landau level to overlap, an electron photoexcited into the lowest spin sublevel of the first Landau level rapidly relaxes into an unoccupied state of the lower-spin sublevel of the zeroth Landau level. Avoiding this process imposes stringent requirements on electron mobility. Even at record-high mobility, however, an electron bound in a CSFE may relax rapidly by another route. The electron in the first Landau level lies above the Fermi level, whereas the Fermi hole lies below it; the Fermi level is set by dopants in the heterostructure barrier. The CSFE lifetime can therefore be limited by tunneling from the excited state into the barrier. Symmetric doping on both sides of the QW, as shown in Fig.~\ref{fig:fig1}(a), suppresses this process. It makes the envelope wave function of the lowest conduction-band size-quantization subband symmetric and reduces the penetration of its tails into the barrier relative to a one-side-doped QW. Moreover, because dopants are present on both sides, the doping layers can be placed farther into the barrier while maintaining the same electron density, further increasing the tunneling time. An additional consequence of symmetric doping is that the growth-direction components of the wave functions of a conduction-band Fermi hole and a valence-band hole are nearly identical. The electron-electron and electron-hole interactions are therefore also nearly equal, which substantially simplifies the description of CSFE-ensemble formation and of the interaction between a CSFE and a valence-band hole \cite{bychkov1991}.

\section{Experimental Results}

Figs.~\ref{fig:fig1}(b) and \ref{fig:fig2} show representative PL spectra collected from two excitation spots whose areas differ by four orders of magnitude, at comparable maximum total pump powers. The spectrum obtained with the large spot [Fig.~\ref{fig:fig1}(b)] contains the plasmaron band Pln and the trion line T, but no magnetoplasmon-molecule lines. The plasmaron band is dominated by the Pln$_M$ line, whose maximum corresponds to the CSFE dispersion minimum at generalized momentum $q\approx1/l_B$, and the Pln$_0$ line, which corresponds to CSFEs with $q\approx0$. With the small excitation spot, additional lines appear alongside the trion and plasmaron features; we attribute these lines to the hybrid magnetoplasmon molecule Mpm (Fig.~\ref{fig:fig2}). We use three experimental criteria to identify this state:

\begin{figure*}[t]
\includegraphics[width=\textwidth]{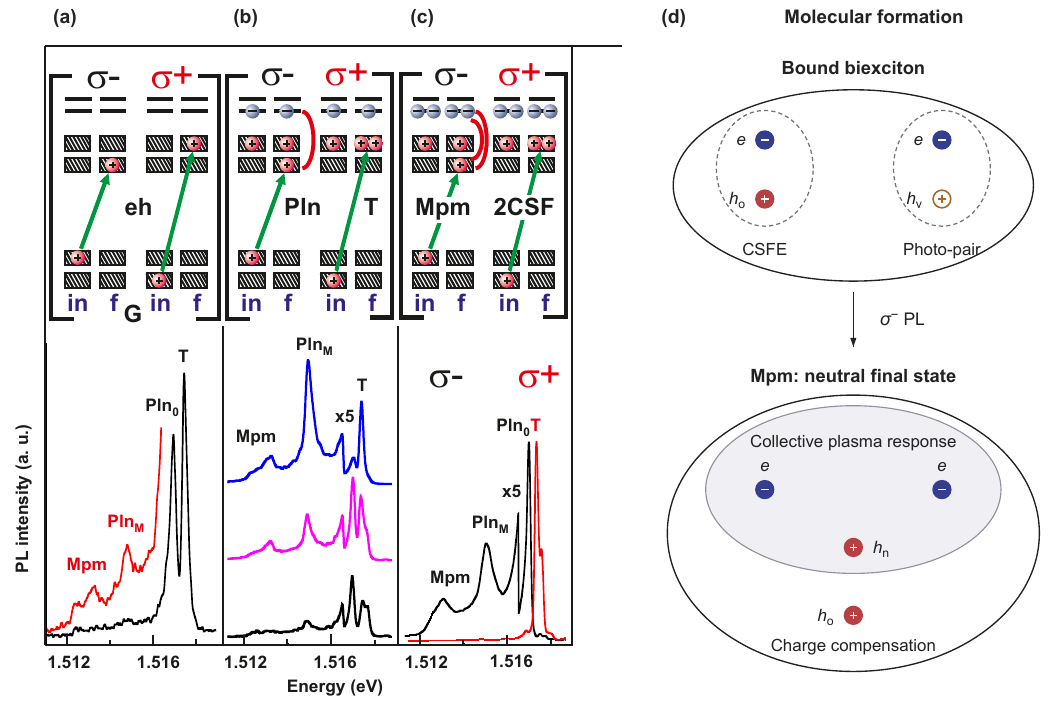}
\caption{PL spectra of the 2DES at $T=600~\mathrm{mK}$ and $d_{\mathrm{pump}}\approx10~\mu\mathrm{m}$. (a,b) Evolution over $I_{\mathrm{pump}}\approx0.1$--$20~\mathrm{W\,cm^{-2}}$. (c) Polarization-resolved molecular (Mpm), plasmaron (Pln$_0$, Pln$_M$), and trion (T) features; the upper diagrams show the initial and final configurations. (d) Schematic molecular channel. Recombination of $h_v$ creates $h_n$; both excited electrons and $h_n$ form the plasma-active component, while $h_o$ compensates its charge. The contours indicate constituent grouping, not a calculated geometry.}
\label{fig:fig2}
\end{figure*}

(i) a four-particle state comprising a CSFE and a photoexcited electron-hole pair must form at the same spatial location; (ii) the molecular lines must occur in the same polarization configuration as the plasmaron; and (iii) the spectrum must contain lines whose final-state energy loss exceeds the maximum plasma-oscillation energy of a single magnetoplasmon.

The first criterion is established from the dependence of the PL intensity on pump power. This dependence must be superlinear, reflecting the increasing probability that the photoexcited valence-band hole and photoexcited electron occupy the same spatial region in the initial state of the recombination process. The second criterion fixes the spin of the valence-band hole before recombination. It ensures that the valence-band hole is converted into a Fermi hole with the same spin as the electron in the CSFE and, consequently, that plasma oscillations are generated in the final state [Fig.~\ref{fig:fig2}(c)]. The third criterion implies that both the photoexcited electron and the electron originally bound in the CSFE simultaneously participate in plasma oscillations with the newly created Fermi hole. Thus, after recombination of the valence-band hole, the CSFE and the photoexcited electron-hole pair remain bound in a single four-particle molecule. This third criterion is sufficient but not necessary: molecules with lower plasma-oscillation energies may also form, but their spectral signatures overlap the plasmaron band and cannot be distinguished from three-particle excitations.

\begin{figure*}[t]
\includegraphics[width=\textwidth]{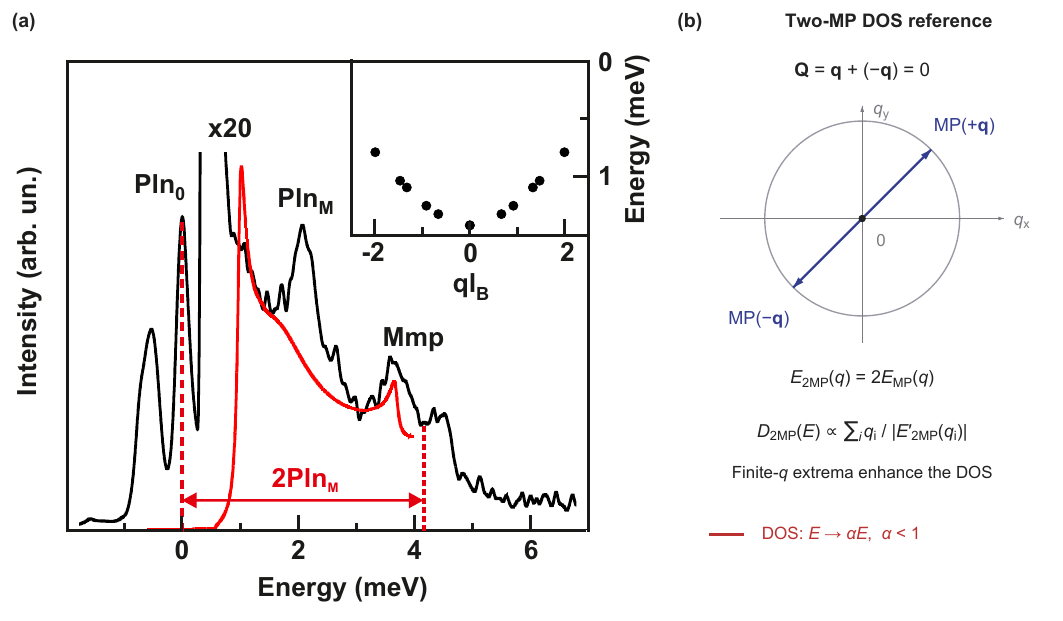}
\caption{(a) Energy-loss spectrum at $T=600~\mathrm{mK}$, $d_{\mathrm{pump}}\approx10~\mu\mathrm{m}$, and $I_{\mathrm{pump}}\approx0.3~\mathrm{W\,cm^{-2}}$ (black). The red curve is the two-magnetoplasmon density of states (DOS) at zero total generalized momentum, rescaled by $\alpha<1$; the inset shows the initial biexciton dispersion. (b) Momentum-space reference: $+q$ and $-q$ give $Q=0$, with $E_{2\mathrm{MP}}(q)=2E_{\mathrm{MP}}(q)$. Finite-$q$ stationary points enhance the DOS. The schematic is not a molecular geometry or an assignment of the two molecular lines to separate van Hove features.}
\label{fig:fig3}
\end{figure*}

\begin{figure*}[t]
\includegraphics[width=\textwidth]{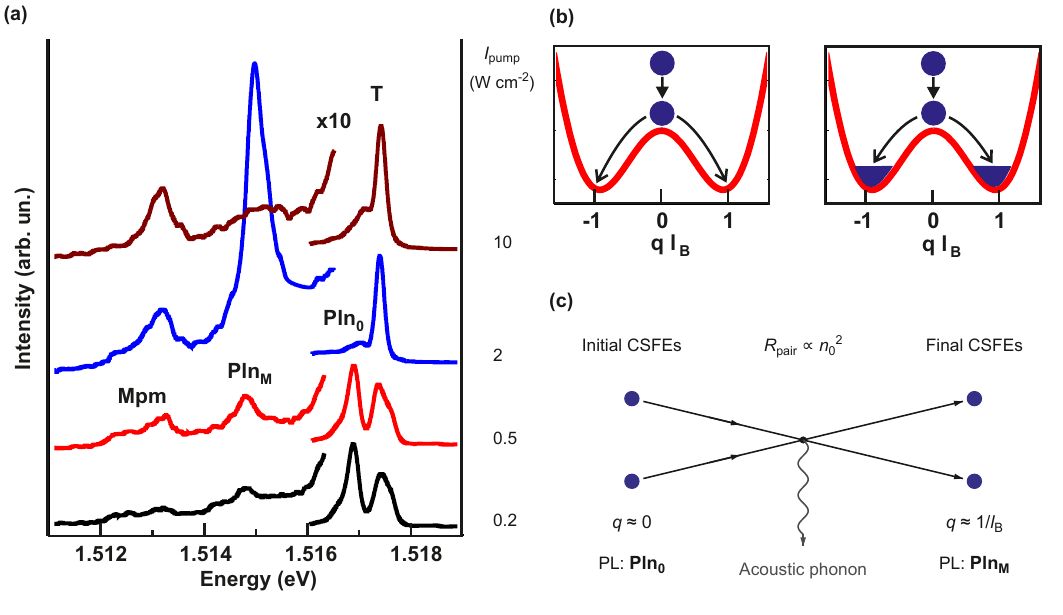}
\caption{(a) Pump-dependent PL as the CSFE ensemble crosses over from gas to condensate at $T=600~\mathrm{mK}$ and $d_{\mathrm{pump}}\approx10~\mu\mathrm{m}$; $I_{\mathrm{pump}}=0.2$, 0.5, 2, and $10~\mathrm{W\,cm^{-2}}$ from bottom to top. (b) Single-particle and collective occupation of the CSFE dispersion minimum. (c) Two-CSFE phonon-assisted relaxation from $q\approx0$ to $q\approx1/l_B$. In the dilute limit, the pair-event rate satisfies $R_{\mathrm{pair}}\propto n_0^2$.}
\label{fig:fig4}
\end{figure*}

\begin{figure*}[t]
\includegraphics[width=\textwidth]{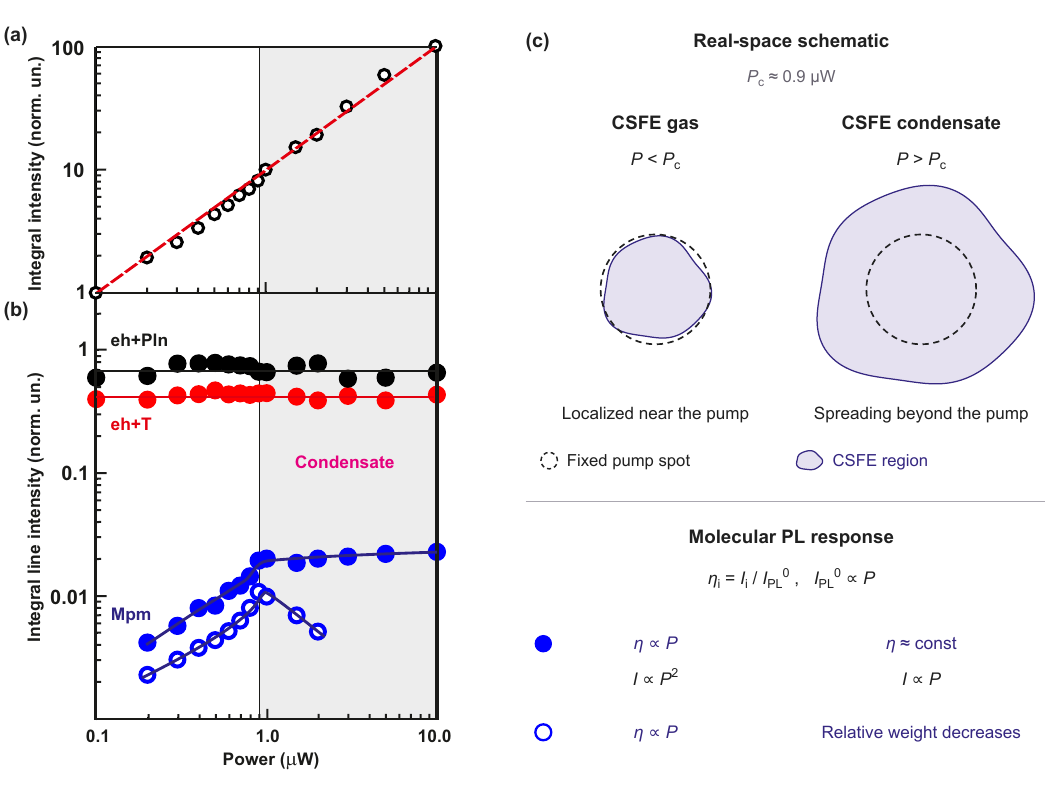}
\caption{(a) Integrated PL intensity versus pump power $P$. (b) PL-channel intensities normalized to the integrated PL; the gray region marks the condensate regime above $P_c\approx0.9~\mu\mathrm{W}$. (c) Qualitative real-space schematic. The pump spot is fixed, whereas the CSFE region spreads in the condensate. Since $\eta_i=I_i/I_{\mathrm{PL}}^0$ and $I_{\mathrm{PL}}^0\propto P$, $\eta_i\propto P$ corresponds to $I_i\propto P^2$, while constant $\eta_i$ corresponds to $I_i\propto P$. The spatial profiles are schematic.}
\label{fig:fig5}
\end{figure*}

The PL energy-loss spectrum contains two lines well above the maximum plasma-oscillation energy, thereby satisfying criterion (iii). Both occur in the $\sigma^-$ polarization configuration, satisfying criterion (ii). One line lies 0.5 meV below, and the other 0.35 meV above, twice the maximum plasmaron energy (Fig.~\ref{fig:fig3}). The PL intensities of the two lines exhibit the same dependence on pump power density. When the CSFE ensemble is in the gas phase, both depend quadratically on pump power (Figs.~\ref{fig:fig4} and \ref{fig:fig5}). Criterion (i) is therefore also satisfied, allowing both high-energy lines to be assigned to the formation of a hybrid magnetoplasmon molecule in the electron system.

Over nearly the entire pump-power range investigated, the quasi-equilibrium CSFE density is high enough for a photoexcited electron-hole pair to form a bound state with a CSFE. At low pump power densities, most CSFEs have zero generalized momentum. As the exciton density increases, occupation of the CSFE dispersion minimum begins abruptly when exciton-exciton scattering becomes the dominant scattering mechanism (Fig.~\ref{fig:fig4}). Below this threshold, individual CSFEs relax to the ground state before acquiring a large in-plane momentum through interactions with heterostructure phonons. Exciton-exciton scattering greatly accelerates momentum relaxation because two zero-momentum excitons can scatter simultaneously into states with large, opposite momenta (Fig.~\ref{fig:fig4}) \cite{dickmann2021}.

\section{Discussion}

A hybrid magnetoplasmon molecule can form only through an interaction between a CSFE and a photoexcited electron-hole pair. No additional conservation laws are required for its formation because the molecule is electrically neutral. The generalized momenta of the initial and final states must nevertheless be conserved. The collective mode selected in the final state is therefore determined not by the momentum of the CSFE alone, as in plasmaron formation, but by conservation of the generalized momentum of the entire four-particle state during valence-hole recombination. Because the emitted photon momentum is negligible relative to the characteristic momenta of the many-particle states, it may be set to zero, and the optical transitions may be treated as vertical in momentum space.

We calculated the ground-state dispersion of the bound four-particle complex (biexciton) in the initial state of recombination, assuming equal electron-electron and electron-hole interactions. The dispersions of single- and multiexciton collective excitations were obtained by exact diagonalization of the energy spectrum for a finite number of electrons at filling factor $\nu=2$. An adequate description of the Coulomb correlations requires multiexciton corrections in both the ground and excited states. The single-mode approximation \cite{longo1993} is therefore inapplicable to both the cyclotron spin-flip dispersion and the molecular states considered here.

The energy spectrum was calculated for electrons on a torus with periodic boundary conditions and a rectangular unit cell. The number of magnetic-flux quanta through the system ranged from $N_S=10$ to $N_S=15$, and the number of electrons at $\nu=2$ was twice as large. States in the two lowest, spin-split Landau levels were included. The Coulomb interaction, with Fourier component $V(q)=[2\pi e^2/(\varepsilon q)]F_{\mathrm{QW}}(q)$, was modified by the geometric form factor $F_{\mathrm{QW}}(q)$, which accounts for the distribution of the electron wave functions along the QW growth direction. These wave functions were obtained from a self-consistent solution of the Schrödinger-Poisson equations. The biexciton spectrum was calculated in a basis of many-electron configurations with fixed total spin projection along the magnetic field, $S_z=+1$. The minimum step in dimensionless momentum was $\delta q=\sqrt{2\pi/N_S}\sim0.65$. Smooth dispersion curves, required to calculate the magnetoplasmon and CSFE densities of states, were obtained from the discrete spectra by ninth-order polynomial interpolation [Fig.~\ref{fig:fig1}(c)]. Fits of the calculated densities of states to the experimental spectra in Figs.~\ref{fig:fig1}(d) and \ref{fig:fig3} used a level-broadening parameter of $\gamma\sim0.01$--$0.02~\mathrm{meV}$.

The biexciton ground state has a monotonic dispersion with a minimum at zero generalized momentum. The measured energies of the magnetoplasmon-molecule lines do not depend on the momentum of the initial CSFE, suggesting that most biexcitons relax to their lowest, zero-momentum states before recombination. The Coulomb binding energy of the biexciton in the initial state should not affect the optical-transition energy because the same binding energy remains in the final state. This conclusion is supported by the optical transitions of two- and three-particle states. For example, although the trion binding energy can be calculated analytically \cite{kuznetsov2018}, it does not enter the optical-transition energy at any initial CSFE momentum. The situation is analogous for a plasmaron formed from a zero-momentum CSFE.

Because molecular momentum is conserved during valence-hole recombination, the possible plasma-oscillation energies of the hybrid magnetoplasmon molecule are described by the density of states of two magnetoplasmons with zero total momentum. Hybrid magnetoplasmon molecules might therefore be expected over the entire interval from zero to twice the maximum magnetoplasmon energy (Fig.~\ref{fig:fig3}). The collective-mode energy of the molecule should, however, be lower than twice the energy of an isolated magnetoplasmon because two electrons cannot interact simultaneously with a single Fermi hole. Their interactions must be ordered in time, reducing the plasma-oscillation energy. Coulomb interaction with the additional Fermi hole, which does not participate in the plasma oscillations, may further reduce this energy, as observed for a plasmaron.

The lower-energy hybrid-molecule line occurs at approximately 0.9 times twice the maximum plasmaron energy. The two-magnetoplasmon density of states contains critical points associated with van Hove singularities, which should produce maxima in the spectrum of hybrid four-particle molecular excitations. One maximum overlaps the three-particle excitation band; we therefore associate the observed characteristic molecular lines with the second maximum. The origin of the two distinct molecular lines remains unclear. The higher-energy line may be related to the truncation of the high-energy portion of the magnetoplasmon dispersion in real electron systems reported in Ref.~\onlinecite{budanov2025}. Consistent with this interpretation, no final states are observed in the PL spectrum above the second molecular line.

The interaction of a magnetoplasmon molecule with the CSFE condensate that forms at $T\lesssim1~\mathrm{K}$ \cite{kulik2016} is nontrivial. Describing the formation dynamics of the CSFE condensate and its interaction with many-particle excitations lies beyond the single-molecule treatment adopted here; instead, we present an experimental analysis of this interaction (Fig.~\ref{fig:fig5}). One established property of the CSFE condensate is a limiting density that cannot be exceeded within the excitation spot \cite{koreyev2024}. This limit arises because every CSFE must contain a well-defined Fermi hole and excited electron, a condition satisfied only when the CSFE density does not exceed approximately one tenth of the equilibrium-electron density in a single spin-resolved Landau sublevel. Otherwise, the Fermi hole is no longer well defined. Excess gas-phase excitons relax to the ground state, whereas excess condensate excitons spread beyond the excitation spot. Before a magnetoexciton condensate forms, the PL intensities of both magnetoplasmon-molecule states increase quadratically with pump power, as expected (Fig.~\ref{fig:fig5}). After condensation, however, the recombination intensity of a photoexcited biexciton into a molecular state becomes linear in pump power. Thus, spatially separated photoexcited electrons and holes find one another within the CSFE condensate with a probability independent of their density. The lower-energy magnetoplasmon-molecule state remains a well-defined four-particle excitation in both the gas phase and the CSFE condensate, whereas recombination into the higher-energy molecular state disappears in the condensate (Figs.~\ref{fig:fig4} and \ref{fig:fig5}).

\section{Conclusion}

We have observed an intrinsic bound four-particle excitation of a 2DES containing a quasi-equilibrium ensemble of long-lived magnetoexcitons: a hybrid magnetoplasmon molecule. The molecular state comprises two identical electrons and two Fermi holes with opposite spins. Both electrons and one Fermi hole participate in the plasma oscillations, whereas the second Fermi hole screens the excess electronic charge without participating in those oscillations. Two lines separated by 0.85 meV in the energy-loss spectrum of valence-hole recombination are associated with the molecular state. We have examined the evolution of the molecular spectrum as the magnetoexciton ensemble crosses over from the gas phase to the condensate.

\begin{acknowledgments}
The authors thank V. D. Kulakovskii for valuable discussions. This work was supported by the Russian Science Foundation (Project No. 24-12-00411).
\end{acknowledgments}

\bibliography{references}

%apsrev4-2.bst 2019-01-14 (MD) hand-edited version of apsrev4-1.bst
%Control: key (0)
%Control: author (8) initials jnrlst
%Control: editor formatted (1) identically to author
%Control: production of article title (0) allowed
%Control: page (0) single
%Control: year (1) truncated
%Control: production of eprint (0) enabled
\begin{thebibliography}{23}%
\makeatletter
\providecommand \@ifxundefined [1]{%
 \@ifx{#1\undefined}
}%
\providecommand \@ifnum [1]{%
 \ifnum #1\expandafter \@firstoftwo
 \else \expandafter \@secondoftwo
 \fi
}%
\providecommand \@ifx [1]{%
 \ifx #1\expandafter \@firstoftwo
 \else \expandafter \@secondoftwo
 \fi
}%
\providecommand \natexlab [1]{#1}%
\providecommand \enquote  [1]{``#1''}%
\providecommand \bibnamefont  [1]{#1}%
\providecommand \bibfnamefont [1]{#1}%
\providecommand \citenamefont [1]{#1}%
\providecommand \href@noop [0]{\@secondoftwo}%
\providecommand \href [0]{\begingroup \@sanitize@url \@href}%
\providecommand \@href[1]{\@@startlink{#1}\@@href}%
\providecommand \@@href[1]{\endgroup#1\@@endlink}%
\providecommand \@sanitize@url [0]{\catcode `\\12\catcode `\$12\catcode
  `\&12\catcode `\#12\catcode `\^12\catcode `\_12\catcode `\%12\relax}%
\providecommand \@@startlink[1]{}%
\providecommand \@@endlink[0]{}%
\providecommand \url  [0]{\begingroup\@sanitize@url \@url }%
\providecommand \@url [1]{\endgroup\@href {#1}{\urlprefix }}%
\providecommand \urlprefix  [0]{URL }%
\providecommand \Eprint [0]{\href }%
\providecommand \doibase [0]{https://doi.org/}%
\providecommand \selectlanguage [0]{\@gobble}%
\providecommand \bibinfo  [0]{\@secondoftwo}%
\providecommand \bibfield  [0]{\@secondoftwo}%
\providecommand \translation [1]{[#1]}%
\providecommand \BibitemOpen [0]{}%
\providecommand \bibitemStop [0]{}%
\providecommand \bibitemNoStop [0]{.\EOS\space}%
\providecommand \EOS [0]{\spacefactor3000\relax}%
\providecommand \BibitemShut  [1]{\csname bibitem#1\endcsname}%
\let\auto@bib@innerbib\@empty
%</preamble>
\bibitem [{\citenamefont {DuBois}\ and\ \citenamefont
  {Goldman}(1978)}]{dubois1978}%
  \BibitemOpen
  \bibfield  {author} {\bibinfo {author} {\bibfnamefont {D.~F.}\ \bibnamefont
  {DuBois}}\ and\ \bibinfo {author} {\bibfnamefont {M.~V.}\ \bibnamefont
  {Goldman}},\ }\bibfield  {title} {\bibinfo {title} {Plasmon-plasmon
  interactions},\ }\href {https://doi.org/10.1103/PhysRevLett.40.257}
  {\bibfield  {journal} {\bibinfo  {journal} {Phys. Rev. Lett.}\ }\textbf
  {\bibinfo {volume} {40}},\ \bibinfo {pages} {257} (\bibinfo {year}
  {1978})}\BibitemShut {NoStop}%
\bibitem [{\citenamefont {Bostwick}\ \emph {et~al.}(2010)\citenamefont
  {Bostwick}, \citenamefont {Speck}, \citenamefont {Seyller}, \citenamefont
  {Horn}, \citenamefont {Polini}, \citenamefont {Asgari}, \citenamefont
  {MacDonald},\ and\ \citenamefont {Rotenberg}}]{bostwick2010}%
  \BibitemOpen
  \bibfield  {author} {\bibinfo {author} {\bibfnamefont {A.}~\bibnamefont
  {Bostwick}}, \bibinfo {author} {\bibfnamefont {F.}~\bibnamefont {Speck}},
  \bibinfo {author} {\bibfnamefont {T.}~\bibnamefont {Seyller}}, \bibinfo
  {author} {\bibfnamefont {K.}~\bibnamefont {Horn}}, \bibinfo {author}
  {\bibfnamefont {M.}~\bibnamefont {Polini}}, \bibinfo {author} {\bibfnamefont
  {R.}~\bibnamefont {Asgari}}, \bibinfo {author} {\bibfnamefont {A.~H.}\
  \bibnamefont {MacDonald}},\ and\ \bibinfo {author} {\bibfnamefont
  {E.}~\bibnamefont {Rotenberg}},\ }\bibfield  {title} {\bibinfo {title}
  {Observation of plasmarons in quasi-freestanding doped graphene},\ }\href
  {https://doi.org/10.1126/science.1186489} {\bibfield  {journal} {\bibinfo
  {journal} {Science}\ }\textbf {\bibinfo {volume} {328}},\ \bibinfo {pages}
  {999} (\bibinfo {year} {2010})}\BibitemShut {NoStop}%
\bibitem [{\citenamefont {Zhuravlev}\ \emph {et~al.}(2016)\citenamefont
  {Zhuravlev}, \citenamefont {Kuznetsov}, \citenamefont {Kulik}, \citenamefont
  {Bisti}, \citenamefont {Kirpichev}, \citenamefont {Kukushkin},\ and\
  \citenamefont {Schmult}}]{zhuravlev2016}%
  \BibitemOpen
  \bibfield  {author} {\bibinfo {author} {\bibfnamefont {A.~S.}\ \bibnamefont
  {Zhuravlev}}, \bibinfo {author} {\bibfnamefont {V.~A.}\ \bibnamefont
  {Kuznetsov}}, \bibinfo {author} {\bibfnamefont {L.~V.}\ \bibnamefont
  {Kulik}}, \bibinfo {author} {\bibfnamefont {V.~E.}\ \bibnamefont {Bisti}},
  \bibinfo {author} {\bibfnamefont {V.~E.}\ \bibnamefont {Kirpichev}}, \bibinfo
  {author} {\bibfnamefont {I.~V.}\ \bibnamefont {Kukushkin}},\ and\ \bibinfo
  {author} {\bibfnamefont {S.}~\bibnamefont {Schmult}},\ }\bibfield  {title}
  {\bibinfo {title} {Artificially constructed plasmarons and plasmon-exciton
  molecules in {2D} metals},\ }\href
  {https://doi.org/10.1103/PhysRevLett.117.196802} {\bibfield  {journal}
  {\bibinfo  {journal} {Phys. Rev. Lett.}\ }\textbf {\bibinfo {volume} {117}},\
  \bibinfo {pages} {196802} (\bibinfo {year} {2016})}\BibitemShut {NoStop}%
\bibitem [{\citenamefont {Dial}\ \emph {et~al.}(2012)\citenamefont {Dial},
  \citenamefont {Ashoori}, \citenamefont {Pfeiffer},\ and\ \citenamefont
  {West}}]{dial2012}%
  \BibitemOpen
  \bibfield  {author} {\bibinfo {author} {\bibfnamefont {O.~E.}\ \bibnamefont
  {Dial}}, \bibinfo {author} {\bibfnamefont {R.~C.}\ \bibnamefont {Ashoori}},
  \bibinfo {author} {\bibfnamefont {L.~N.}\ \bibnamefont {Pfeiffer}},\ and\
  \bibinfo {author} {\bibfnamefont {K.~W.}\ \bibnamefont {West}},\ }\bibfield
  {title} {\bibinfo {title} {Observations of plasmarons in a two-dimensional
  system: Tunneling measurements using time-domain capacitance spectroscopy},\
  }\href {https://doi.org/10.1103/PhysRevB.85.081306} {\bibfield  {journal}
  {\bibinfo  {journal} {Phys. Rev. B}\ }\textbf {\bibinfo {volume} {85}},\
  \bibinfo {pages} {081306(R)} (\bibinfo {year} {2012})}\BibitemShut {NoStop}%
\bibitem [{\citenamefont {Fofang}\ \emph {et~al.}(2011)\citenamefont {Fofang},
  \citenamefont {Grady}, \citenamefont {Fan}, \citenamefont {Govorov},\ and\
  \citenamefont {Halas}}]{fofang2011}%
  \BibitemOpen
  \bibfield  {author} {\bibinfo {author} {\bibfnamefont {N.~T.}\ \bibnamefont
  {Fofang}}, \bibinfo {author} {\bibfnamefont {N.~K.}\ \bibnamefont {Grady}},
  \bibinfo {author} {\bibfnamefont {Z.}~\bibnamefont {Fan}}, \bibinfo {author}
  {\bibfnamefont {A.~O.}\ \bibnamefont {Govorov}},\ and\ \bibinfo {author}
  {\bibfnamefont {N.~J.}\ \bibnamefont {Halas}},\ }\bibfield  {title} {\bibinfo
  {title} {Plexciton dynamics: Exciton--plasmon coupling in a
  {J}-aggregate--{Au} nanoshell complex provides a mechanism for
  nonlinearity},\ }\href {https://doi.org/10.1021/nl104352j} {\bibfield
  {journal} {\bibinfo  {journal} {Nano Lett.}\ }\textbf {\bibinfo {volume}
  {11}},\ \bibinfo {pages} {1556} (\bibinfo {year} {2011})}\BibitemShut
  {NoStop}%
\bibitem [{\citenamefont {Balci}\ and\ \citenamefont
  {Kocabas}(2015)}]{balci2015}%
  \BibitemOpen
  \bibfield  {author} {\bibinfo {author} {\bibfnamefont {S.}~\bibnamefont
  {Balci}}\ and\ \bibinfo {author} {\bibfnamefont {C.}~\bibnamefont
  {Kocabas}},\ }\bibfield  {title} {\bibinfo {title} {Ultra hybrid plasmonics:
  Strong coupling of plexcitons with plasmon polaritons},\ }\href
  {https://doi.org/10.1364/OL.40.003424} {\bibfield  {journal} {\bibinfo
  {journal} {Opt. Lett.}\ }\textbf {\bibinfo {volume} {40}},\ \bibinfo {pages}
  {3424} (\bibinfo {year} {2015})}\BibitemShut {NoStop}%
\bibitem [{\citenamefont {T{\"o}rm{\"a}}\ and\ \citenamefont
  {Barnes}(2015)}]{torma2015}%
  \BibitemOpen
  \bibfield  {author} {\bibinfo {author} {\bibfnamefont {P.}~\bibnamefont
  {T{\"o}rm{\"a}}}\ and\ \bibinfo {author} {\bibfnamefont {W.~L.}\ \bibnamefont
  {Barnes}},\ }\bibfield  {title} {\bibinfo {title} {Strong coupling between
  surface plasmon polaritons and emitters: A review},\ }\href
  {https://doi.org/10.1088/0034-4885/78/1/013901} {\bibfield  {journal}
  {\bibinfo  {journal} {Rep. Prog. Phys.}\ }\textbf {\bibinfo {volume} {78}},\
  \bibinfo {pages} {013901} (\bibinfo {year} {2015})}\BibitemShut {NoStop}%
\bibitem [{\citenamefont {Zhao}\ \emph {et~al.}(2025)\citenamefont {Zhao},
  \citenamefont {L{\"u}}, \citenamefont {Sun},\ and\ \citenamefont
  {Wu}}]{zhao2025}%
  \BibitemOpen
  \bibfield  {author} {\bibinfo {author} {\bibfnamefont {P.}~\bibnamefont
  {Zhao}}, \bibinfo {author} {\bibfnamefont {C.}~\bibnamefont {L{\"u}}},
  \bibinfo {author} {\bibfnamefont {S.}~\bibnamefont {Sun}},\ and\ \bibinfo
  {author} {\bibfnamefont {F.}~\bibnamefont {Wu}},\ }\bibfield  {title}
  {\bibinfo {title} {Plasmon--exciton strong coupling in low-dimensional
  materials: From fundamentals to hybrid nanophotonic platforms},\ }\href
  {https://doi.org/10.3390/nano15191463} {\bibfield  {journal} {\bibinfo
  {journal} {Nanomaterials}\ }\textbf {\bibinfo {volume} {15}},\ \bibinfo
  {pages} {1463} (\bibinfo {year} {2025})}\BibitemShut {NoStop}%
\bibitem [{\citenamefont {Kulik}\ \emph {et~al.}(2000)\citenamefont {Kulik},
  \citenamefont {Kukushkin}, \citenamefont {Kirpichev}, \citenamefont {von
  Klitzing},\ and\ \citenamefont {Eberl}}]{kulik2000}%
  \BibitemOpen
  \bibfield  {author} {\bibinfo {author} {\bibfnamefont {L.~V.}\ \bibnamefont
  {Kulik}}, \bibinfo {author} {\bibfnamefont {I.~V.}\ \bibnamefont
  {Kukushkin}}, \bibinfo {author} {\bibfnamefont {V.~E.}\ \bibnamefont
  {Kirpichev}}, \bibinfo {author} {\bibfnamefont {K.}~\bibnamefont {von
  Klitzing}},\ and\ \bibinfo {author} {\bibfnamefont {K.}~\bibnamefont
  {Eberl}},\ }\bibfield  {title} {\bibinfo {title} {Interaction between
  intersubband {Bernstein} modes and coupled plasmon-phonon modes},\ }\href
  {https://doi.org/10.1103/PhysRevB.61.12717} {\bibfield  {journal} {\bibinfo
  {journal} {Phys. Rev. B}\ }\textbf {\bibinfo {volume} {61}},\ \bibinfo
  {pages} {12717} (\bibinfo {year} {2000})}\BibitemShut {NoStop}%
\bibitem [{\citenamefont {Bychkov}\ \emph {et~al.}(1981)\citenamefont
  {Bychkov}, \citenamefont {Iordanskii},\ and\ \citenamefont
  {Eliashberg}}]{bychkov1981}%
  \BibitemOpen
  \bibfield  {author} {\bibinfo {author} {\bibfnamefont {Y.~A.}\ \bibnamefont
  {Bychkov}}, \bibinfo {author} {\bibfnamefont {S.~V.}\ \bibnamefont
  {Iordanskii}},\ and\ \bibinfo {author} {\bibfnamefont {G.~M.}\ \bibnamefont
  {Eliashberg}},\ }\bibfield  {title} {\bibinfo {title} {Two-dimensional
  electrons in a strong magnetic field},\ }\href@noop {} {\bibfield  {journal}
  {\bibinfo  {journal} {JETP Lett.}\ }\textbf {\bibinfo {volume} {33}},\
  \bibinfo {pages} {143} (\bibinfo {year} {1981})}\BibitemShut {NoStop}%
\bibitem [{\citenamefont {Kallin}\ and\ \citenamefont
  {Halperin}(1984)}]{kallin1984}%
  \BibitemOpen
  \bibfield  {author} {\bibinfo {author} {\bibfnamefont {C.}~\bibnamefont
  {Kallin}}\ and\ \bibinfo {author} {\bibfnamefont {B.~I.}\ \bibnamefont
  {Halperin}},\ }\bibfield  {title} {\bibinfo {title} {Excitations from a
  filled {Landau} level in the two-dimensional electron gas},\ }\href
  {https://doi.org/10.1103/PhysRevB.30.5655} {\bibfield  {journal} {\bibinfo
  {journal} {Phys. Rev. B}\ }\textbf {\bibinfo {volume} {30}},\ \bibinfo
  {pages} {5655} (\bibinfo {year} {1984})}\BibitemShut {NoStop}%
\bibitem [{\citenamefont {Kulik}\ \emph {et~al.}(2005)\citenamefont {Kulik},
  \citenamefont {Kukushkin}, \citenamefont {Dickmann}, \citenamefont
  {Kirpichev}, \citenamefont {Van'kov}, \citenamefont {Parakhonsky},
  \citenamefont {Smet}, \citenamefont {von Klitzing},\ and\ \citenamefont
  {Wegscheider}}]{kulik2005}%
  \BibitemOpen
  \bibfield  {author} {\bibinfo {author} {\bibfnamefont {L.~V.}\ \bibnamefont
  {Kulik}}, \bibinfo {author} {\bibfnamefont {I.~V.}\ \bibnamefont
  {Kukushkin}}, \bibinfo {author} {\bibfnamefont {S.}~\bibnamefont {Dickmann}},
  \bibinfo {author} {\bibfnamefont {V.~E.}\ \bibnamefont {Kirpichev}}, \bibinfo
  {author} {\bibfnamefont {A.~B.}\ \bibnamefont {Van'kov}}, \bibinfo {author}
  {\bibfnamefont {A.~L.}\ \bibnamefont {Parakhonsky}}, \bibinfo {author}
  {\bibfnamefont {J.~H.}\ \bibnamefont {Smet}}, \bibinfo {author}
  {\bibfnamefont {K.}~\bibnamefont {von Klitzing}},\ and\ \bibinfo {author}
  {\bibfnamefont {W.}~\bibnamefont {Wegscheider}},\ }\bibfield  {title}
  {\bibinfo {title} {Cyclotron spin-flip mode as the lowest-energy excitation
  of unpolarized integer quantum {Hall} states},\ }\href
  {https://doi.org/10.1103/PhysRevB.72.073304} {\bibfield  {journal} {\bibinfo
  {journal} {Phys. Rev. B}\ }\textbf {\bibinfo {volume} {72}},\ \bibinfo
  {pages} {073304} (\bibinfo {year} {2005})}\BibitemShut {NoStop}%
\bibitem [{\citenamefont {Kulik}\ \emph {et~al.}(2015)\citenamefont {Kulik},
  \citenamefont {Gorbunov}, \citenamefont {Zhuravlev}, \citenamefont
  {Timofeev}, \citenamefont {Dickmann},\ and\ \citenamefont
  {Kukushkin}}]{kulik2015}%
  \BibitemOpen
  \bibfield  {author} {\bibinfo {author} {\bibfnamefont {L.~V.}\ \bibnamefont
  {Kulik}}, \bibinfo {author} {\bibfnamefont {A.~V.}\ \bibnamefont {Gorbunov}},
  \bibinfo {author} {\bibfnamefont {A.~S.}\ \bibnamefont {Zhuravlev}}, \bibinfo
  {author} {\bibfnamefont {V.~B.}\ \bibnamefont {Timofeev}}, \bibinfo {author}
  {\bibfnamefont {S.}~\bibnamefont {Dickmann}},\ and\ \bibinfo {author}
  {\bibfnamefont {I.~V.}\ \bibnamefont {Kukushkin}},\ }\bibfield  {title}
  {\bibinfo {title} {Super-long life time for {2D} cyclotron spin-flip
  excitons},\ }\href {https://doi.org/10.1038/srep10354} {\bibfield  {journal}
  {\bibinfo  {journal} {Sci. Rep.}\ }\textbf {\bibinfo {volume} {5}},\ \bibinfo
  {pages} {10354} (\bibinfo {year} {2015})}\BibitemShut {NoStop}%
\bibitem [{\citenamefont {Dickmann}(2013)}]{dickmann2013}%
  \BibitemOpen
  \bibfield  {author} {\bibinfo {author} {\bibfnamefont {S.}~\bibnamefont
  {Dickmann}},\ }\bibfield  {title} {\bibinfo {title} {Extremely slow spin
  relaxation in a spin-unpolarized quantum {Hall} system},\ }\href
  {https://doi.org/10.1103/PhysRevLett.110.166801} {\bibfield  {journal}
  {\bibinfo  {journal} {Phys. Rev. Lett.}\ }\textbf {\bibinfo {volume} {110}},\
  \bibinfo {pages} {166801} (\bibinfo {year} {2013})}\BibitemShut {NoStop}%
\bibitem [{\citenamefont {Kulik}\ \emph {et~al.}(2016)\citenamefont {Kulik},
  \citenamefont {Zhuravlev}, \citenamefont {Dickmann}, \citenamefont
  {Gorbunov}, \citenamefont {Timofeev}, \citenamefont {Kukushkin},\ and\
  \citenamefont {Schmult}}]{kulik2016}%
  \BibitemOpen
  \bibfield  {author} {\bibinfo {author} {\bibfnamefont {L.~V.}\ \bibnamefont
  {Kulik}}, \bibinfo {author} {\bibfnamefont {A.~S.}\ \bibnamefont
  {Zhuravlev}}, \bibinfo {author} {\bibfnamefont {S.}~\bibnamefont {Dickmann}},
  \bibinfo {author} {\bibfnamefont {A.~V.}\ \bibnamefont {Gorbunov}}, \bibinfo
  {author} {\bibfnamefont {V.~B.}\ \bibnamefont {Timofeev}}, \bibinfo {author}
  {\bibfnamefont {I.~V.}\ \bibnamefont {Kukushkin}},\ and\ \bibinfo {author}
  {\bibfnamefont {S.}~\bibnamefont {Schmult}},\ }\bibfield  {title} {\bibinfo
  {title} {Magnetofermionic condensate in two dimensions},\ }\href
  {https://doi.org/10.1038/ncomms13499} {\bibfield  {journal} {\bibinfo
  {journal} {Nat. Commun.}\ }\textbf {\bibinfo {volume} {7}},\ \bibinfo {pages}
  {13499} (\bibinfo {year} {2016})}\BibitemShut {NoStop}%
\bibitem [{\citenamefont {Kuznetsov}\ \emph {et~al.}(2018)\citenamefont
  {Kuznetsov}, \citenamefont {Kulik}, \citenamefont {Velikanov}, \citenamefont
  {Zhuravlev}, \citenamefont {Gorbunov}, \citenamefont {Schmult},\ and\
  \citenamefont {Kukushkin}}]{kuznetsov2018}%
  \BibitemOpen
  \bibfield  {author} {\bibinfo {author} {\bibfnamefont {V.~A.}\ \bibnamefont
  {Kuznetsov}}, \bibinfo {author} {\bibfnamefont {L.~V.}\ \bibnamefont
  {Kulik}}, \bibinfo {author} {\bibfnamefont {M.~D.}\ \bibnamefont
  {Velikanov}}, \bibinfo {author} {\bibfnamefont {A.~S.}\ \bibnamefont
  {Zhuravlev}}, \bibinfo {author} {\bibfnamefont {A.~V.}\ \bibnamefont
  {Gorbunov}}, \bibinfo {author} {\bibfnamefont {S.}~\bibnamefont {Schmult}},\
  and\ \bibinfo {author} {\bibfnamefont {I.~V.}\ \bibnamefont {Kukushkin}},\
  }\bibfield  {title} {\bibinfo {title} {Three-particle electron-hole complexes
  in two-dimensional electron systems},\ }\href
  {https://doi.org/10.1103/PhysRevB.98.205303} {\bibfield  {journal} {\bibinfo
  {journal} {Phys. Rev. B}\ }\textbf {\bibinfo {volume} {98}},\ \bibinfo
  {pages} {205303} (\bibinfo {year} {2018})}\BibitemShut {NoStop}%
\bibitem [{\citenamefont {Dickmann}\ \emph {et~al.}(2019)\citenamefont
  {Dickmann}, \citenamefont {Kulik},\ and\ \citenamefont
  {Kuznetsov}}]{dickmann2019}%
  \BibitemOpen
  \bibfield  {author} {\bibinfo {author} {\bibfnamefont {S.}~\bibnamefont
  {Dickmann}}, \bibinfo {author} {\bibfnamefont {L.~V.}\ \bibnamefont
  {Kulik}},\ and\ \bibinfo {author} {\bibfnamefont {V.~A.}\ \bibnamefont
  {Kuznetsov}},\ }\bibfield  {title} {\bibinfo {title} {Coherence-decoherence
  transition in a spin-magnetoexcitonic ensemble in a quantum {Hall} system},\
  }\href {https://doi.org/10.1103/PhysRevB.100.155304} {\bibfield  {journal}
  {\bibinfo  {journal} {Phys. Rev. B}\ }\textbf {\bibinfo {volume} {100}},\
  \bibinfo {pages} {155304} (\bibinfo {year} {2019})}\BibitemShut {NoStop}%
\bibitem [{\citenamefont {Avron}\ \emph {et~al.}(1978)\citenamefont {Avron},
  \citenamefont {Herbst},\ and\ \citenamefont {Simon}}]{avron1978}%
  \BibitemOpen
  \bibfield  {author} {\bibinfo {author} {\bibfnamefont {J.~E.}\ \bibnamefont
  {Avron}}, \bibinfo {author} {\bibfnamefont {I.~W.}\ \bibnamefont {Herbst}},\
  and\ \bibinfo {author} {\bibfnamefont {B.}~\bibnamefont {Simon}},\ }\bibfield
   {title} {\bibinfo {title} {Separation of center of mass in homogeneous
  magnetic fields},\ }\href {https://doi.org/10.1016/0003-4916(78)90276-2}
  {\bibfield  {journal} {\bibinfo  {journal} {Ann. Phys. (N.Y.)}\ }\textbf
  {\bibinfo {volume} {114}},\ \bibinfo {pages} {431} (\bibinfo {year}
  {1978})}\BibitemShut {NoStop}%
\bibitem [{\citenamefont {Bychkov}\ and\ \citenamefont
  {Rashba}(1991)}]{bychkov1991}%
  \BibitemOpen
  \bibfield  {author} {\bibinfo {author} {\bibfnamefont {Y.~A.}\ \bibnamefont
  {Bychkov}}\ and\ \bibinfo {author} {\bibfnamefont {E.~I.}\ \bibnamefont
  {Rashba}},\ }\bibfield  {title} {\bibinfo {title} {Excitons and deexcitons in
  a neutral two-dimensional magnetoplasma with a strong population inversion},\
  }\href {https://doi.org/10.1103/PhysRevB.44.6212} {\bibfield  {journal}
  {\bibinfo  {journal} {Phys. Rev. B}\ }\textbf {\bibinfo {volume} {44}},\
  \bibinfo {pages} {6212} (\bibinfo {year} {1991})}\BibitemShut {NoStop}%
\bibitem [{\citenamefont {Dickmann}\ and\ \citenamefont
  {Kaysin}(2021)}]{dickmann2021}%
  \BibitemOpen
  \bibfield  {author} {\bibinfo {author} {\bibfnamefont {S.}~\bibnamefont
  {Dickmann}}\ and\ \bibinfo {author} {\bibfnamefont {B.~D.}\ \bibnamefont
  {Kaysin}},\ }\bibfield  {title} {\bibinfo {title} {Stochastization of long
  living spin-cyclotron excitations in a spin-unpolarised quantum {Hall}
  system},\ }\href {https://doi.org/10.1134/S002136402122001X} {\bibfield
  {journal} {\bibinfo  {journal} {JETP Lett.}\ }\textbf {\bibinfo {volume}
  {114}},\ \bibinfo {pages} {585} (\bibinfo {year} {2021})}\BibitemShut
  {NoStop}%
\bibitem [{\citenamefont {Longo}\ and\ \citenamefont
  {Kallin}(1993)}]{longo1993}%
  \BibitemOpen
  \bibfield  {author} {\bibinfo {author} {\bibfnamefont {J.~P.}\ \bibnamefont
  {Longo}}\ and\ \bibinfo {author} {\bibfnamefont {C.}~\bibnamefont {Kallin}},\
  }\bibfield  {title} {\bibinfo {title} {Spin-flip excitations from {Landau}
  levels in two dimensions},\ }\href {https://doi.org/10.1103/PhysRevB.47.4429}
  {\bibfield  {journal} {\bibinfo  {journal} {Phys. Rev. B}\ }\textbf {\bibinfo
  {volume} {47}},\ \bibinfo {pages} {4429} (\bibinfo {year}
  {1993})}\BibitemShut {NoStop}%
\bibitem [{\citenamefont {Budanov}\ \emph {et~al.}(2025)\citenamefont
  {Budanov}, \citenamefont {Shchigarev}, \citenamefont {Kulik}, \citenamefont
  {Van'kov}, \citenamefont {Larionov}, \citenamefont {Gorbunov},\ and\
  \citenamefont {Umansky}}]{budanov2025}%
  \BibitemOpen
  \bibfield  {author} {\bibinfo {author} {\bibfnamefont {E.~M.}\ \bibnamefont
  {Budanov}}, \bibinfo {author} {\bibfnamefont {D.~A.}\ \bibnamefont
  {Shchigarev}}, \bibinfo {author} {\bibfnamefont {L.~V.}\ \bibnamefont
  {Kulik}}, \bibinfo {author} {\bibfnamefont {A.~B.}\ \bibnamefont {Van'kov}},
  \bibinfo {author} {\bibfnamefont {A.~V.}\ \bibnamefont {Larionov}}, \bibinfo
  {author} {\bibfnamefont {A.~V.}\ \bibnamefont {Gorbunov}},\ and\ \bibinfo
  {author} {\bibfnamefont {V.}~\bibnamefont {Umansky}},\ }\bibfield  {title}
  {\bibinfo {title} {Measurement of the magnetoplasmon density of states in the
  radiative recombination spectrum of symmetrically doped quantum wells},\
  }\href {https://doi.org/10.1134/S0021364025605871} {\bibfield  {journal}
  {\bibinfo  {journal} {JETP Lett.}\ }\textbf {\bibinfo {volume} {121}},\
  \bibinfo {pages} {648} (\bibinfo {year} {2025})}\BibitemShut {NoStop}%
\bibitem [{\citenamefont {Koreyev}\ \emph {et~al.}(2024)\citenamefont
  {Koreyev}, \citenamefont {Berezhnoy}, \citenamefont {Gorbunov}, \citenamefont
  {Solovyev}, \citenamefont {Van'kov}, \citenamefont {Kulik},\ and\
  \citenamefont {Timofeev}}]{koreyev2024}%
  \BibitemOpen
  \bibfield  {author} {\bibinfo {author} {\bibfnamefont {A.~S.}\ \bibnamefont
  {Koreyev}}, \bibinfo {author} {\bibfnamefont {P.~S.}\ \bibnamefont
  {Berezhnoy}}, \bibinfo {author} {\bibfnamefont {A.~V.}\ \bibnamefont
  {Gorbunov}}, \bibinfo {author} {\bibfnamefont {V.~V.}\ \bibnamefont
  {Solovyev}}, \bibinfo {author} {\bibfnamefont {A.~B.}\ \bibnamefont
  {Van'kov}}, \bibinfo {author} {\bibfnamefont {L.~V.}\ \bibnamefont {Kulik}},\
  and\ \bibinfo {author} {\bibfnamefont {V.~B.}\ \bibnamefont {Timofeev}},\
  }\bibfield  {title} {\bibinfo {title} {Phase diagram of magnetoexciton
  condensate},\ }\href {https://doi.org/10.1103/PhysRevB.110.165417} {\bibfield
   {journal} {\bibinfo  {journal} {Phys. Rev. B}\ }\textbf {\bibinfo {volume}
  {110}},\ \bibinfo {pages} {165417} (\bibinfo {year} {2024})}\BibitemShut
  {NoStop}%
\end{thebibliography}%

\end{document}